\documentclass[aps,prl,twocolumn,showpacs,superscriptaddress,groupedaddress]{revtex4}  

\usepackage{color, units, graphicx, amsmath, amssymb, bm}
\usepackage{braket}
\usepackage{epstopdf}
\usepackage{natbib}
\usepackage{changes}

\makeatletter
\newcommand{\rmnum}[1]{\romannumeral #1}
\newcommand{\Rmnum}[1]{\expandafter\@slowromancap\romannumeral #1@}
\makeatother

\include{macros}

\begin{document}

\title{Geometrical pumping with a Bose-Einstein condensate}

\author{H.-I Lu}
\affiliation{Joint Quantum Institute, National Institute of Standards and Technology, and University of Maryland, Gaithersburg, Maryland, 20899, USA}
\author{M. Schemmer}
\affiliation{Joint Quantum Institute, National Institute of Standards and Technology, and University of Maryland, Gaithersburg, Maryland, 20899, USA}
\affiliation{\'{E}cole Normale Sup\'{e}rieure de Lyon, F-69364 Lyon, France}
\author{L. M. Aycock}
\affiliation{Joint Quantum Institute, National Institute of Standards and Technology, and University of Maryland, Gaithersburg, Maryland, 20899, USA}
\affiliation{Cornell University, Ithaca, New York, 14850, USA}
\author{D. Genkina}
\affiliation{Joint Quantum Institute, National Institute of Standards and Technology, and University of Maryland, Gaithersburg, Maryland, 20899, USA}
\author{S. Sugawa}
\affiliation{Joint Quantum Institute, National Institute of Standards and Technology, and University of Maryland, Gaithersburg, Maryland, 20899, USA}
\author{I. B. Spielman}
\email{ian.spielman@nist.gov}
\affiliation{Joint Quantum Institute, National Institute of Standards and Technology, and University of Maryland, Gaithersburg, Maryland, 20899, USA}

\date{\today}

\begin{abstract}
We realized a quantum geometric ``charge'' pump for a Bose-Einstein condensate (BEC) in the lowest Bloch band of a novel bipartite magnetic lattice. Topological charge pumps in filled bands yield quantized pumping set by the global -- topological -- properties of the bands. In contrast, our geometric charge pump for a BEC occupying just a single crystal momentum state  exibits non-quantized charge pumping set by local -- geometrical -- properties of the band structure. Like topological charge pumps, for each pump cycle we observed an overall displacement (here, not quantized) and a temporal modulation of the atomic wavepacket's position in each unit cell, i.e., the polarization.
\end{abstract}
\maketitle

Ultracold atoms in optical lattices provide a unique setting for experimentally studying concepts that lie at the heart of theoretical condensed matter physics, but are out of reach of current condensed matter experiments. Here we focus on the connection between topology, geometry, and adiabatic charge pumping~\cite{Mei2014,Zhang2015,Wang2013,Wei2015,Zeng2015,Marra2015,Cooper2015} for Bose-Einstein condensates (BECs) in cyclically driven lattice potentials.

Particles in periodic potentials form Bloch bands with energy $\epsilon_n(q)$ and eigenstates $\ket{\Psi_n(q)} = \exp(i q \hat x) \ket{u_n(q)}$ labeled by the crystal momentum $q$ along with the band index $n$. The states $\ket{u_n}$ retain the underlying periodicity of the lattice, set by the unit cell size $a$. Motion in lattices is conventionally understood in terms of these bands: metals are materials with partially filled bands, while insulators have completely filled bands. In this context, a topological charge pump is a counterintuitive device, where charge motion -- conduction -- accompanies the adiabatic and cyclic drive of an insulating lattice's parameters. Thouless showed that this conduction is quantized, completely governed by the band-topology~\cite{Thouless1983,Niu1984}. Although various charge pumps have been realized in condensed matter devices -- such as modulated quantum dots~\cite{Switkes1999,Blumenthal2007,Kaestner2008}, 1D channels driven by surface acoustic waves~\cite{Talyanskii1997}, and superconducting qubits~\cite{Mottonen2008} -- Thouless pumps remain unrealized in condensed matter settings but have been demonstrated in recent experiments with cold-atom insulators~\cite{Nakajima2015,Loshe2015}.

Here we break from this established paradigm for insulators and create a quantum charge pump for a BEC in a one dimensional (1D) lattice~\cite{Yaschenko1998,Resta2000,Qian2011} occupying a single crystal momentum state $q$. This charge pump gives non-quantized motion sensitive to the Berry curvature at $q$ integrated over the whole pump cycle, a local geometric quantity, rather than a global topological quantity. Berry curvatures play an important role in condensed matter systems. An iconic example is the integer quantum Hall effect, where the electrons acquire an anomalous transverse velocity proportional to the Berry curvature and the quantized Hall conductance is given by the Berry curvature integrated over the whole 2D Brillouin zone (BZ)~\cite{Thouless1982}; recent cold-atom experiments in 2D have measured such curvatures integrated over part~\cite{Jotzu2014,Duca2015} or all~\cite{Aidelsburger2015} of the BZ. In an analogous way, 1D lattice systems, driven cyclically in time $t$, have a generalized Berry curvature defined on the 2D effective BZ in $q,t$ space. This curvature is the source of an anomalous velocity~\cite{Shen2012}, utilized to drive an adiabatic quantum pumping process.

The Rice-Mele model~\cite{Su1979,Rice1982,Xiao2010,Atala2013} of a bipartite lattice with a unit cell consisting of $A$ and $B$ sites is the paradigmatic system for understanding quantum pumps. The Hamiltonian for this tight-binding model is
\begin{align}
\hat H_{\rm RM} =& -\sum_j \left[\left(t+\delta t\right)\hat b^\dagger_j \hat a_j + \left(t-\delta t\right)\hat a^\dagger_{j+1} \hat b_{j} + {\rm H.c} \right] \nonumber\\
&+ \Delta\sum_j \left(\hat a^\dagger_{j} \hat a_{j} - \hat b^\dagger_{j} \hat b_{j} \right),
\end{align}
where $\hat a^\dagger_j$ and $\hat b^\dagger_j$ describe the creation of a particle in unit cell $j$ and sublattice site $A$ or $B$ respectively. The nominal tunneling strength $t$ is staggered by $\delta t$, and the sublattice sites are shifted in energy by $\Delta$.

\begin{figure}[t!]
  \centering
    \includegraphics{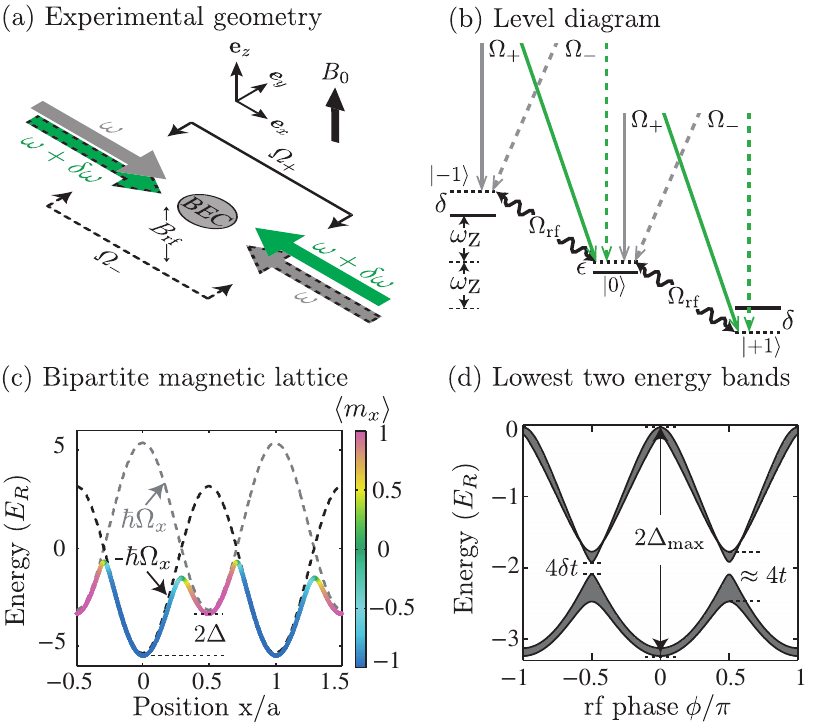}
    \caption{Bipartite magnetic lattice. (a,b) Dipole trapped $^{87}{\rm Rb}$ BECs subject to a bias magnetic field $B_0{\textbf e}_z$ had a Zeeman splitting $\omega_{\rm Z}/2\pi= 0.817\ {\rm MHz}$ and a quadratic shift $\hbar\epsilon=0.03E_R$. These BECs were illuminated by four Raman beams and an rf magnetic field. Each of the two Raman couplings (strengths $\Omega_{\pm}$) was derived from two cross-polarized Raman laser beams with frequency components $\omega$ and $\omega+\delta \omega$. (c) Adiabatic potentials colored according to $\langle m_x \rangle$ computed for $\hbar({\bar{\Omega},\Omega_{\rm rf}},\delta)=(6,2.2,0)E_R$, $\delta\Omega/\bar\Omega=-0.1$, and $\phi = \pi/4$. The dashed curves plot the $\pm\hbar\Omega_x$ contributions to the potential experienced by states $\ket{m_x=\pm1}$. (d) Lowest two energy band energies plotted as a function of $\phi$, otherwise with the same parameters as (c).
    }
    \label{Setup}
\end{figure}

We investigated quantum pumping in a novel 1D (along ${\bf e}_x$) bipartite magnetic lattice (building on Refs.~\cite{Jimenez-Garcia2012,Cheuk2012}) that in effect allowed independent control of $t$, $\delta t$, and $\Delta$.  As shown in Fig.~\ref{Setup}(a)-(b), our magnetic lattice for $^{87}{\rm Rb}$ arose from the interplay of one rf and two ``Raman'' fields that coupled the $\ket{f=1; m_F = \pm1,0}$ ``spin'' states comprising the $f=1$ ground state hyperfine manifold, which were Zeeman split by $\hbar\omega_Z$. The natural units of momentum and energy are given by the single photon recoil momentum $\hbar k_{R}\!=\!2\pi\hbar/\lambda_{R}$ and its corresponding energy $E_{R}\!=\!\hbar^2k_{R}^2/2m$, where $m$ is the atomic mass. In the frame rotating at the rf frequency $\delta\omega$ and under the rotating wave approximation, the combined rf/Raman coupling lead~\cite{Juzeliunas2012} to the overall Hamiltonian
\begin{equation}
 \hat{H}=\frac{\hbar^2\hat k_x^2}{2m}+\mathbf{\Omega}( \hat{x}) \cdot  \hat{\mathbf{F}}+ \hat{H}_Q,
\label{eq:Hamiltonian}
\end{equation}
where $\hat{\mathbf{F}}$ is the total angular momentum vector operator. We interpret $\mathbf{\Omega}(\hat x)=[\Omega_{\rm rf}\cos(\phi)+\bar{\Omega}\cos(2k_R\hat x),-\Omega_{\rm rf}\sin(\phi)-\delta\Omega \sin(2k_R\hat x),\sqrt{2}\delta]/\sqrt{2}$ as a spatially periodic effective Zeeman magnetic field, in which: $\Omega_{\rm rf}$ is the rf coupling strength; $\bar{\Omega}=\Omega_+ +\Omega_-$ and $\delta\Omega=\Omega_+ -\Omega_-$ are derived from the individual Raman coupling strengths $\Omega_\pm$; $\delta\! = \!\delta\omega\!-\! \omega_Z\!$~is the detuning from Raman/rf resonance; $\phi$ is the relative phase between the rf and Raman fields. Additionally, $H_{\rm Q}\!\!=\!\!-\epsilon(\hbar^2 \hat {\mathbb{I}}\!\! -\!\! \hat{F}_{z}^2)/\hbar$ describes the quadratic Zeeman shift, where $\hat{\mathbb{I}}$ is the identity operator.

This spatially varying effective magnetic field produces a 1D bipartite lattice~\cite{Lundblad2014,Zhang2015} with lattice constant $a=\lambda_R/2$ with adiabatic (Born-Oppenheimer) potentials depicted in Fig.~\ref{Setup}(c). This magnetic lattice is most easily conceptualized for small $\delta\Omega$: the $\bar{\Omega}\cos(2k_R \hat x)$ term provides periodic potentials for the $\ket{m_x=\pm1}$ states spatially displaced from each other by $a/2$ [dashed curves in Fig.~\ref{Setup}(c)]; the resulting $m_x = \pm1$ sites are then staggered in energy, giving $\Delta\approx\Delta_{\rm max}\cos(\phi)$, with $\Delta_{\rm max} = \Omega_{\rm rf}/\sqrt 2$.  The $\Omega_y$ term couples these sublattices together: the rf term $-\Omega_{\rm rf}\sin(\phi)$ generates constant height barriers (largely specifying $t$), which become staggered by the $-\delta\Omega \sin(2k_R\hat x)$ contribution (largely specifying $\delta t$).

Figure~\ref{Setup}(d) plots the energies of the resulting lowest two bands as a function of $\phi$ (modulating $\Delta$ cosinusoidally). Although our lattice is not in the tight binding limit, the band structure qualitatively matches that of the Rice-Mele model. In the remainder of this article, we focus on the lowest band $n=0$ and will henceforth omit the band index.

\begin{figure}[!tb]
  \centering
    \includegraphics{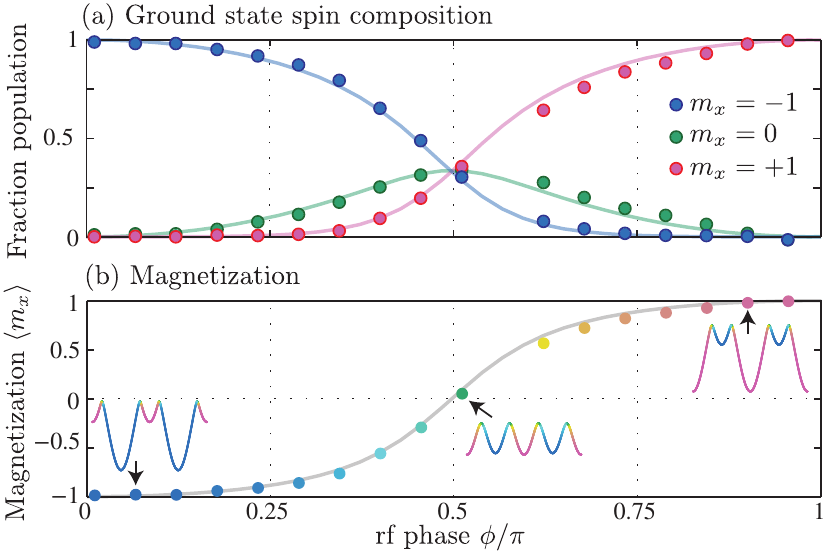}
    \caption{Ground state spin projections. (a) Ground state spin projections at various $\phi$ along with the predicted populations for $\hbar(\bar{\Omega},\delta\Omega,\Omega_{\rm rf},\delta)=(4.4,0,2.2,0)E_R$. The associated adiabatic potentials [insets in (b)] have minima with spin projection following the observed population's trends. (b) Magnetization derived from data in (a).
    }
    \label{StatePrep}
\end{figure}

As illustrated by the shading in Fig.~\ref{Setup}(c), in each unit cell the sublattice sites are ``labeled'' by their $\hat F_x$ spin projection with $\ket{m_x = -1}$ site on the left and $\ket{m_x = +1}$ site on the right.  To confirm this, we adiabatically loaded $\ket{m_z = 0}$ BECs into the lattice's ground state by simultaneously ramping the detuning from $5E_R$ to zero while ramping on the coupling fields in 10 ms. Following preparation, our measurement sequence began with a $\pi/2$ spin rotation along ${\textbf e}_y$, allowing us to measure the eigenstates of $\hat F_x$ in our $\hat F_z$ measurement basis. We achieved this $\pi/2$ rotation with a $44\ \mu{\rm s}$ pulse from an additional rf field with phase $\phi_{\rm rot}=\pi/2$ and strength $\hbar \Omega_{\rm rf, \rm rot}=2.2E_R$, applied while the Raman coupling was greatly reduced ($\bar\Omega\ll\Omega_{\rm rf, \rm rot}$) and the lattice rf coupling was off ($\Omega_{\rm rf} = 0$). We then abruptly removed the remaining control fields along with the confining potential and absorption imaged the resulting spin-resolved momentum distribution after a $20\ {\rm ms}$ time-of-flight (TOF) period in the presence of a magnetic field gradient along ${\bf e}_y$.

Figure~\ref{StatePrep} shows the measured $\hat F_x$ spin composition~\endnote{The data presented in Fig.~\ref{StatePrep} contains a small correction from the known imperfect state rotation.} and magnetization for adiabatically loaded BECs as a function of $\phi$ with $\delta\Omega = 0$. Because $\Delta(\phi)$ controls the relative depth of the $\ket{m_x=\pm1}$ wells, we observe ground states spin populations that follow this ``tilt''. For example, when $\phi=0$ or $\pi$ the double-well is strongly tilted and we observe the near perfect spin magnetization, consistent with atoms residing in the individual sub-lattices; in contrast, when $\phi=\pi/2$, the double-wells are balanced and we observe equal populations in each $\ket{m_x}$ state as expected for equal occupancy of both sub-lattices. Thus the magnetization [Fig.~\ref{StatePrep}(b)] measures the mean atomic position within each unit cell, i.e., the polarization.

Having constructed a physical realization of the Rice-Mele model, and demonstrated the requisite control and measurement tools, we now turn our attention to topological and geometrical charge pumping. These fundamentally quantum mechanical effects rely on the canonical commutation relation between position and momentum. Consider a finite wavepacket with center of mass position (COM) $\langle x\rangle=\langle\Psi|\hat x|\Psi\rangle$, subject to a lattice Hamiltonian $\hat H$ that is adiabatically modulated with period $T$, i.e., $\hat H(t)=\hat H(t+T)$. After one cycle, any initial crystal momentum state is transformed
$\ket{\Psi(q)}\rightarrow \exp(i\gamma(\hat q))\ket{\Psi(q)}$, at most acquiring a phase, where $\hat q$ is the crystal momentum operator; this defines the single-period evolution operator $\hat U_T = \exp(i\gamma(\hat q))$. The time-evolved position operator $\hat U^\dagger_T \hat x \hat U_T = \hat x - \partial_{\hat q} \gamma(\hat q) $ is displaced after a single pump cycle.

The displacement is particularly simple in two limits: when just a single crystal momentum state is occupied or when every crystal momentum state in the BZ, $-\pi/a \leq q <\pi/a$, is occupied with equal probability. As for our BEC, when a single $\ket{q_0}$ state is occupied the displacement is \mbox{$\Delta x(q_0) = - \partial_{q} \gamma(q)|_{q_0}$}. Both the dynamical phase $\gamma_{\rm D}(q) = - \bar \epsilon(q) T / \hbar$ from the time-average energy $\bar \epsilon(q)$, and the geometric Berry phase $\gamma_{\rm B}(q) = i\int_0^T \langle u|\partial_t u\rangle dt$ contribute to $\gamma(q) = \gamma_{\rm D}(q) + \gamma_{\rm B}(q)$. In agreement with conventional descriptions~\cite{Resta2000,Xiao2010,Shen2012}, this predicts a mean velocity $\bar v(q)=\partial_q \bar\epsilon(q)/\hbar -T^{-1}\int_0^T F(q,t)dt$. The first term is the usual group velocity and the second term -- the anomalous velocity -- derives from the Berry curvature $F(q,t)=i(\langle \partial_q u| \partial_t u\rangle- \langle\partial_t u|\partial_q u\rangle)$. In our experiment, the BEC occupied the minimum of $\epsilon(q,t)$ at $q=0$ during the whole pump cycle giving $\partial_q\bar\epsilon(q)=0$, so only the geometric phase $\gamma_{\rm B}(q)$ contributed to the per-cycle displacement $\Delta x(q=0)=-\int_0^T  F(q=0,t) dt$.

In the contrasting case of a filled band, the average group velocity is also zero and the displacement is $\Delta x = - a \int_{\rm BZ}\partial_{q} \gamma_{\rm B}(q) dq/2\pi$; this is often expressed as $\Delta x = a \int_0^T \partial_{t} \gamma_{\rm Zak}(t) dt/{\rm 2\pi}$.  The Zak phase $\gamma_{\rm Zak} = i\int_{\rm BZ} \langle u|\partial_q u\rangle dq$, a topological property of 1D bands, is the Berry's phase associated with traversing the 1D BZ once, in the same way that $\gamma_{\rm B}(q)$ is a Berry's phase taken over a pump cycle.

Our lattice's Zak phase is plotted in Fig.~\ref{Potentials}(a); this Zak phase is qualitatively indistinguishable from that of the Rice-Mele model, with singularities at $\phi=\pm \pi/2$ and $\delta\Omega=0$, signaling topological phase transitions across these points. For filled band experiments, pumping trajectories encircling these points give quantized charge pumping~\cite{Nakajima2015,Loshe2015}. Figure~\ref{Potentials}(b) shows the richly structured Berry curvature $F(q=0,\phi)$ relevant to our experiment, which will be explored next.

\begin{figure}[]
  \centering
    \includegraphics{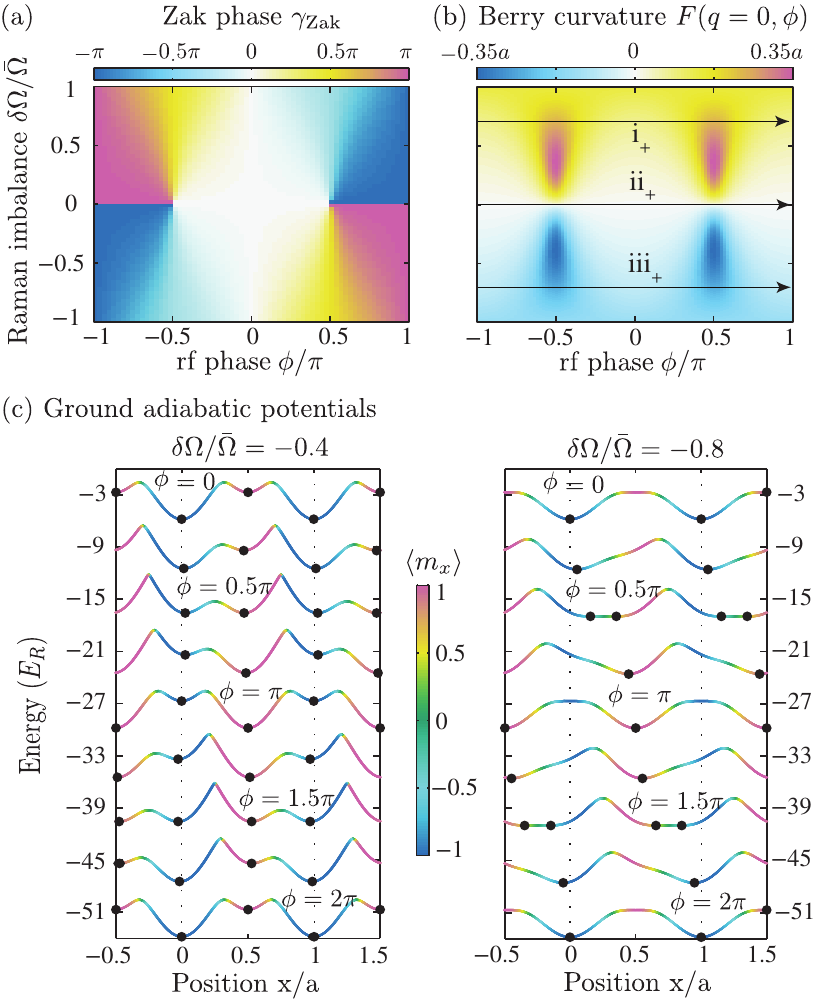}
    \caption{Band geometry and topology computed for $\hbar({\bar{\Omega},\Omega_{\rm rf}},\delta)=(6,2.2,0)E_R$. (a,b) Zak phase and $q=0$ Berry curvature showing the dependence on both $\delta\Omega/\bar\Omega$ and $\phi$. In (b), the arrows show experimental charge pump trajectories in Fig.~\ref{SpinPumpdata}(b). (c) Adiabatic potentials (displaced vertically for clarity) computed for a range of $\phi$ constituting a complete pump cycle at $\delta\Omega/\bar\Omega=-0.4$ (left panel) and $-0.8$ (right panel). Filled circles mark the local energy minima.}
    \label{Potentials}
\end{figure}

For our charge pump experiments, we linearly ramped the pump control parameter $\phi(t)=2\pi t/T$, effectively modulating the lattice potential in two qualitatively different regimes (separated by a critical $|\delta\Omega/\bar\Omega|\approx0.63$).
In the first [Fig.~\ref{Potentials}(c), left panel] the sublattice sites rise and fall but the local potential minima are essentially fixed in space;
in the second [Fig.~\ref{Potentials}(d), right panel] each minimum is only present for part of the pump cycle (the potential appears to ``slide'' by $\pm a$ per cycle).
As these schematics imply, the associated pumping process gives either no displacement, or a quantized per-cycle displacement $\pm a$ for classical trajectories~\cite{SM2015}.
In quantum systems, however, geometrical pumping is controlled by the Berry curvature, giving  non-quantized per-cycle displacements that can in principle take on any value.

\begin{figure*}[]
  \centering
    \includegraphics{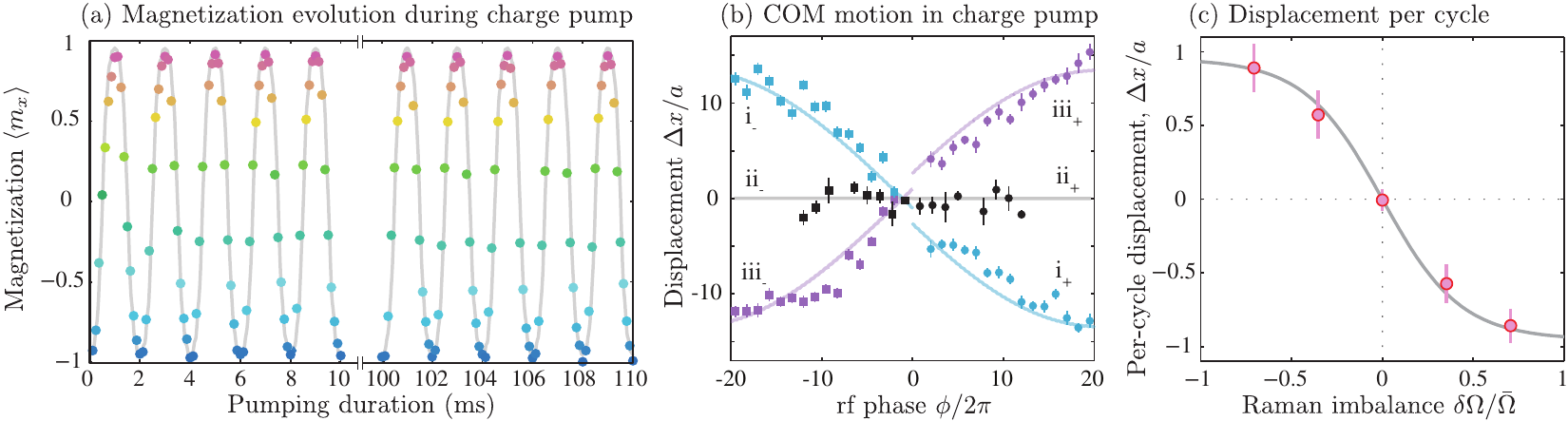}
    \caption{Geometric charge pumping. (a) Magnetization measured while linearly ramping $\phi$ with period $T=2$ ms, along with the prediction for $\hbar\bar{\Omega}=6.38(2)E_R$, $\hbar\delta\Omega=4.50 (2)E_R$, and $\hbar\Omega_{rf}=2.20 (3)E_R$. (b) Displacement plotted versus $\phi/2\pi$ (number of pump cycles).
    Trajectories \rmnum{1}~to \rmnum{3}~are taken at $\delta\Omega/\bar{\Omega}=0.7,0,$ and $-0.7$, respectively; in each case $\hbar\bar{\Omega}\approx6E_R$ and $\hbar\Omega_{rf}=2.20 (3)E_R$. Solid curves: simulation of charge pump in the trap. The small displacement near $\phi=0$ is introduced by our loading procedure. (c) Measured displacement $\Delta x$ per pump cycle (symbols), along with the prediction obtained by integrating the Berry curvature over our pumping trajectory (solid curve). The uncertainty bars represent 95$\%$ confidence interval.
        }
    \label{SpinPumpdata}
\end{figure*}

We studied adiabatic charge pumping in this lattice in two ways: in the first we observed the $\hat F_x$-magnetization, giving the polarization within the unit cells, and in the second we directly measured the displacement $\Delta x$ of our BEC.
In both cases we loaded into the lattice's ground state and linearly ramped $\phi = 2\pi t/T$, driving the Hamiltonian with period $T$~\cite{SM2015}.
As shown in Fig.~\ref{SpinPumpdata}(a), the magnetization oscillated with the $T=2$ ms period, demonstrating the periodic modulation of polarization per cycle.
In good agreement with our data, the solid curves in Fig.~\ref{SpinPumpdata}(a) show the predicted behavior given our known system parameters. This agreement persists to long times: for example after 50 pumping cycles (for $t=100-110$ ms) the contrast is unchanged, confirming the adiabaticity of the process~\cite{SM2015}.

Lastly, we performed a charge pumping experiment by directly measuring the cloud's position in-situ for a range of $\delta\Omega/\bar\Omega$.
We obtained in-situ density distributions using partial-transfer absorption imaging~\cite{Ramanathan2012} in which $\approx 6.8\ {\rm GHz}$ microwave pulses transferred $\approx5\%$ of the atoms from $\ket{f,m_z}=\ket{1,-1}$ to $\ket{2,0}$ where they were absorption imaged. This technique allowed us to repeatedly measure the in-situ density distribution for each BEC.
Each observed displacement was derived from differential measurements of the cloud position taken just before and just after the pumping process, rendering our observations insensitive to micron-level drift in the trap position between different realizations.

Figure~\ref{SpinPumpdata}(b) shows data taken for $\delta\Omega/\bar\Omega=0.7$, $0$, and $-0.7$ along trajectories \rmnum{1}, \rmnum{2}, and \rmnum{3}, respectively, with both increasing and decreasing phases.
Our data displays two expected symmetry properties. First, since the displacement $\Delta x(q=0)=-\int F(q,\phi) d\phi$ depends on the sign of the acquired phase, the direction of motion is reversed when the ramp direction is inverted. Secondly as shown in Fig.~\ref{Potentials}(b), $F(q=0, \phi)$ is an odd function of $\delta\Omega / \bar\Omega$, so the direction of motion is also reversed when $\delta\Omega/\bar\Omega\rightarrow-\delta\Omega/\bar\Omega$. Thus $\Delta x$ is an odd function of both $\phi$ and $\delta\Omega / \bar\Omega$, and as expected we observe no motion when $\delta\Omega/\bar\Omega=0$.

The displacement was markedly non-linear when the pumping time became comparable to our trap's 80 ms period, showing the influence of the confining potential~\endnote{The effective mass in the lattice can increase the period.}.
We included the harmonic potential in our real-space simulations by directly solving the time-dependent Schr\"odinger equation for our system~\cite{Bao2003}.
The simulated results [Fig.~\ref{SpinPumpdata}(b), solid curves] agree with our observations.
To extract the per-cycle displacement due to geometric pumping, we fit the sinusoidal predictions of our model to each data trace, with only the overall amplitudes and a small vertical offset as free parameters, giving the short-time per-cycle displacement~\cite{SM2015}.
Figure~\ref{SpinPumpdata}(c) shows these per-cycle displacements for a range of Raman imbalances.

The in-situ cloud typically had a Thomas-Fermi radius of $30$ $\mu$m, corresponding to a small momentum width of $0.004k_R$ for our BEC. We estimated the thermal fraction to be $\approx 5\%$ given by our $\approx 20$ nK temperature (momentum width of $0.24 k_R$). Moreover, the per-cycle displacement is nearly independent of $q$ for $|q|<0.25 k_R$~\cite{SM2015}. These allow us to compare the data with the expected displacement from integrating $q=0$ Berry curvature [Fig.~\ref{SpinPumpdata}(c), solid line], showing an excellent agreement and confirming the geometric origin of our quantum charge pump.

Our magnetic lattice enables new experiments with 1D topological lattices.
Berry curvatures at $q\neq 0$ can be probed by performing the charge pump pairwise at $\pm|q|$ (for example prepared via Bloch oscillations~\cite{Dahan1996}). The dynamical phases in these cases are opposite and therefore cancel while Berry curvatures (even in $q$) contribute equally to the displacements~\cite{SM2015}.
Furthermore, protected edges states, a hallmark of topological systems, will be present at the interface between regions characterized by different topological invariants~\cite{Ruostekoski2008,Hasan2010,Ganeshan2013}.
Since in our lattice the topological index is set by the rf phase, a bulk topological junction can be generated by replacing the rf field with an additional co-propagating pair of Raman laser beams in which just one beam has an abrupt $\pi$ phase shift in its center. This provides a static model of the soliton excitation mode in polyacetylene~\cite{Su1979,Heeger1988}.
Terminating our lattice with hard-wall boundaries will give rise to similar end states -- somewhat analogous to Majorana fermions in 1D topological superconductors~\cite{Hasan2010,Nadj-perge2014} -- with a spin character.

\begin{acknowledgements}
We appreciate the constructive discussions with W. D. Phillips, E. Mueller, L. J. LeBlanc, and L. Wang.
This work was partially supported by the ARO's Atomtronics MURI, and by the AFOSR's Quantum Matter MURI, NIST, and the NSF through the PFC at the JQI. M.S. was supported by Amp\`{e}re Scholarships of Excellence of the ENS de Lyon. S. S. acknowledges support from JSPS Postdoctoral Fellowship for Research Abroad.
\end{acknowledgements}

\bibliography{GeometricPump}

\pagebreak

\title{Supplementary Material for `Geometrical pumping with a Bose-Einstein condensate'}

\author{H.-I Lu}
\affiliation{Joint Quantum Institute, National Institute of Standards and Technology, and University of Maryland, Gaithersburg, Maryland, 20899, USA}
\author{M. Schemmer}
\affiliation{Joint Quantum Institute, National Institute of Standards and Technology, and University of Maryland, Gaithersburg, Maryland, 20899, USA}
\affiliation{\'{E}cole Normale Sup\'{e}rieure de Lyon, F-69364 Lyon, France}
\author{L. M. Aycock}
\affiliation{Joint Quantum Institute, National Institute of Standards and Technology, and University of Maryland, Gaithersburg, Maryland, 20899, USA}
\affiliation{Cornell University, Ithaca, New York, 14850, USA}
\author{D. Genkina}
\affiliation{Joint Quantum Institute, National Institute of Standards and Technology, and University of Maryland, Gaithersburg, Maryland, 20899, USA}
\author{S. Sugawa}
\affiliation{Joint Quantum Institute, National Institute of Standards and Technology, and University of Maryland, Gaithersburg, Maryland, 20899, USA}
\author{I. B. Spielman}
\email{ian.spielman@nist.gov}
\affiliation{Joint Quantum Institute, National Institute of Standards and Technology, and University of Maryland, Gaithersburg, Maryland, 20899, USA}



\maketitle

\section{Supplementary Material}
\section{Relevant Raman coupling}

\begin{figure*}[]
  \centering
    \includegraphics[width=4.5 in]{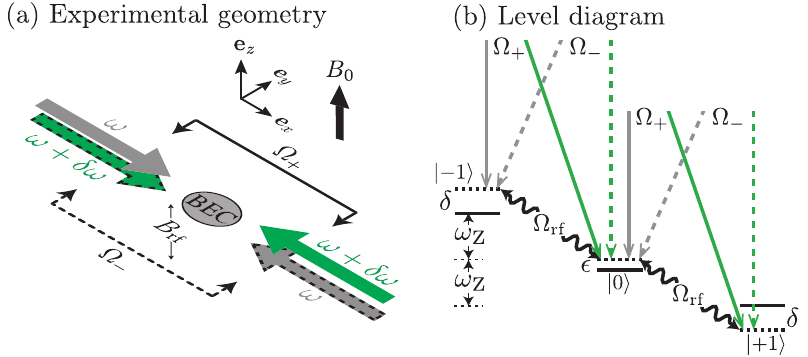}
    \caption{(a) Experimental geometry. (b) Level diagram of magnetic levels dressed by two Raman couplings with strength $\Omega_{\pm}$ and one rf coupling with strength $\Omega_{\rm rf}$.}
    \label{LevelDiagram}
\end{figure*}

The Raman coupling resulting from the vector light shift (proportional to $ \mathbf{E} \times \mathbf{E}$) is non-zero for our cross-polarized counter-propagating lasers, but is zero for the otherwise resonant contributions with each beam taken separately.
For example, two Raman beams propagating along $+{\bf e}_x$ with frequencies $\omega$ and $\omega+\delta\omega$ have the same polarization along ${\bf e}_z$ in Fig.~\ref{LevelDiagram}(a) and hence give no vector light shift. The same frequency components from the cross-polarized Raman beams [e.g., beams with frequency $\omega$ propagating along $+{\bf e}_x$ with polarization along ${\bf e}_z$ and propagating along $-{\bf e}_x$ with polarization along ${\bf e}_y$ in Fig.~\ref{LevelDiagram}(a)] do not provide a time dependent vector light shift for coupling the magnetic levels (or far off resonant from the Zeeman splitting in the rotating wave approximation).
Therefore, the only two relevant Raman coupling pairs are marked in Fig.~\ref{LevelDiagram}(a), with coupling strengths denoted as $\Omega_{\pm}$.

Figure~\ref{LevelDiagram}(b) shows the level diagram of the three $m_F$ states in our $^{87}$Rb BECs $f=1$ ground state manifold as coupled by the experimental combination of Raman and rf coupling.
For example, atoms starting at rest ($k_i=0$) in the $\ket{m_F = -1}$ spin state are coupled to the $\ket{m_F=0}$ state in three ways: the rf field provides coupling but leaves the momentum unchanged at $\hbar k=0$; while the two photon Raman transitions with strength $\Omega_\pm$ change the momentum to $\hbar k=\pm 2 \hbar k_R$.

We independently controlled the amplitudes of $\Omega_{\pm}$ using two rf sources with frequencies $\omega$ and $\omega+\delta\omega$, produced from a Direct Digital Synthesized (DDS) signal generator (Novatech, Model 409B). We generated the desired Raman frequency pairs by splitting the output of each of the rf sources into two and controlled these four amplitudes individually. We identify these four rf signals by $A_+(\omega)$, $A_-(\omega)$, $A_+(\omega+\delta\omega)$, and $A_-(\omega+\delta\omega)$. Here, $A_+$ and $A_-$ denote the rf signal amplitudes. We then combined $A_+(\omega)$ with $A_-(\omega+\delta\omega)$ and used the resulting signal to drive the acousto-optic modulator for the Raman beams propagating along $+{\bf e}_x$ with polarization along ${\bf e}_z$ in Fig.~\ref{LevelDiagram}(a). Similarly we combined $A_+(\omega+\delta\omega)$ and $A_-(\omega)$ to drive the acousto-optic modulator for the Raman beams traveling along $-{\bf e}_x$.

\section{Lattice characterization via pulsing experiment}
\begin{figure*}[]
  \centering
    \includegraphics[width=4.5in]{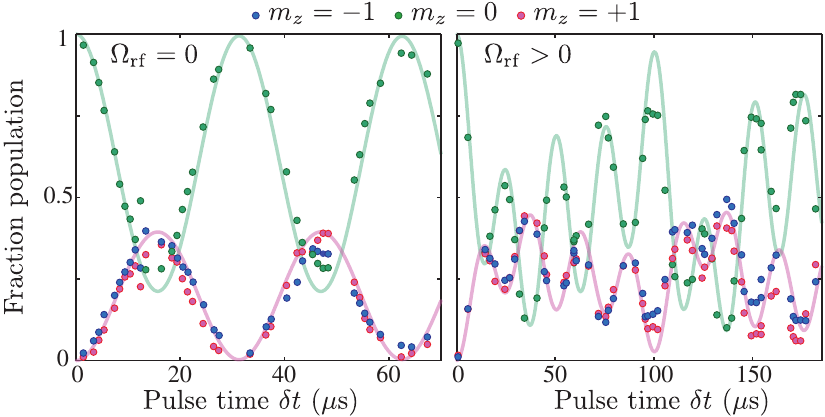}
    \caption{Time evolution of population in $\ket{m_z=0,\pm1}$ states following the instantaneous turn-on of the lattice. Left panel: Simple evolution when only the Raman fields are applied with $\hbar\Omega_+=5.45(2)E_R$ and $\hbar \Omega_-=0.94(1)E_R$. Right panel: Evolution when an rf field with strength $\hbar\Omega_{\rm rf} = 2.20(3) E_R$ and phase $\phi=-0.02(1)\pi$ is added. In each case the solid curves depict the time evolution expected from our Hamiltonian using calibrated couplings; only
    $\phi$ is a fit parameter in the right panel.}
    \label{MakeLattice}
\end{figure*}

We first characterized the magnetic lattice by studying the time-evolution of an initial $\ket{m_z=0}$ BEC following the abrupt turn-on of different combinations of our coupling fields, as shown in Fig.~\ref{MakeLattice}. After holding these fields constant for a time $\delta t$ we abruptly removed them along with the confining potential. We absorption imaged the resulting spin-resolved momentum distribution after a $20\ {\rm ms}$ time-of-flight (TOF) period in the presence of a magnetic field gradient along ${\bf e}_y$. Absent rf coupling we observed sinusoidal Rabi-like oscillations between the $\ket{m_z=0}$ and $\ket{m_z=\pm1}$ states (Fig.~\ref{MakeLattice}, left panel).
While completing the lattice with the rf coupling makes the time-evolution more complex, it is still in good agreement with our predictions (Fig.~\ref{MakeLattice}, right panel). The individual coupling strengths were separately calibrated using a similar technique, and by fitting data in the right panel of Fig.~\ref{MakeLattice} with $\phi$ as the only free parameter, we readily calibrated the rf phase.

\section{Light-atom interaction and control of rf phase} \label{rfphase}
The light-atom interaction due to a two-photon Raman transition can be expressed as
\begin{equation}
 \hat H_{l-a}=\mathbf{\Omega}_{R,\pm} \cdot  \hat{\mathbf{F}}.
\label{eq:H_la}
\end{equation}
The effective magnetic field from a single Raman coupling is
\begin{equation}
 \mathbf{\Omega}_{R,\pm}=\frac{1}{\sqrt2}[\Omega_{\pm}\cos(2k_R\hat x),-\Omega_{\pm}\sin(\pm 2k_R\hat x),\sqrt 2\delta ],
\label{eq:B_Raman}
\end{equation}
where the two coupling fields have an opposite sign on the ${\bf e}_y$ component due to the opposite momentum kick.
The rf magnetic field contributes the additional coupling $\mathbf{\Omega}_{\rm rf}=1/\sqrt2[\Omega_{\rm rf}\cos(\phi),-\Omega_{\rm rf}\sin(\phi),\sqrt 2\delta]$, where the rf phase $\phi$ is defined relative to that of the Raman coupling. The effective magnetic field in Eq.~(2) of the main text has the contributions from the above three coupling fields, $\mathbf{\Omega}_{R,+}$, $\mathbf{\Omega}_{R,-}$, and $\mathbf{\Omega}_{\rm rf}$.
The DDS signal generator described in the previous section has four rf phase synchronous outputs, and we used two of the rf sources to generate frequency shifts on the Raman lasers. In addition, we used the other two rf sources to generate the rf couplings for lattice and $\pi/2$ spin rotation. The path lengths were different for the signal cables which drove the Raman AOMs and the rf coil. In addition, Raman beams after AOMs were sent to the chamber through optical fibers. The phase difference between the Raman couplings and rf coupling controlled by the DDS signal generator was different from the relative phase experienced by atoms. We therefore used both the pulsing and adiabatic loading measurements (Fig.~\ref{MakeLattice} and Fig.~2 in the main text) to calibrate the relative phase on atoms at a certain commanded phase difference. After calibrating the phase shift introduced by the different paths, we can add the known phase shift value to the DDS signal generator.
In our charge pumping experiment, we periodically modulated the lattice potential by linearly ramping the rf phase. To achieve this, the rf frequency was set to $\delta \omega_{\rm rf}=\delta \omega+\delta \omega'$, corresponding to a linearly varying phase relative to that of the Raman fields with $T=2\pi/\delta \omega'$.
We switched between linearly increasing or decreasing phases by inverting the sign of $\delta \omega'$.

\section{Classical, geometric, and topological pumps}
\begin{figure*}
  \centering
    \includegraphics[width=5in]{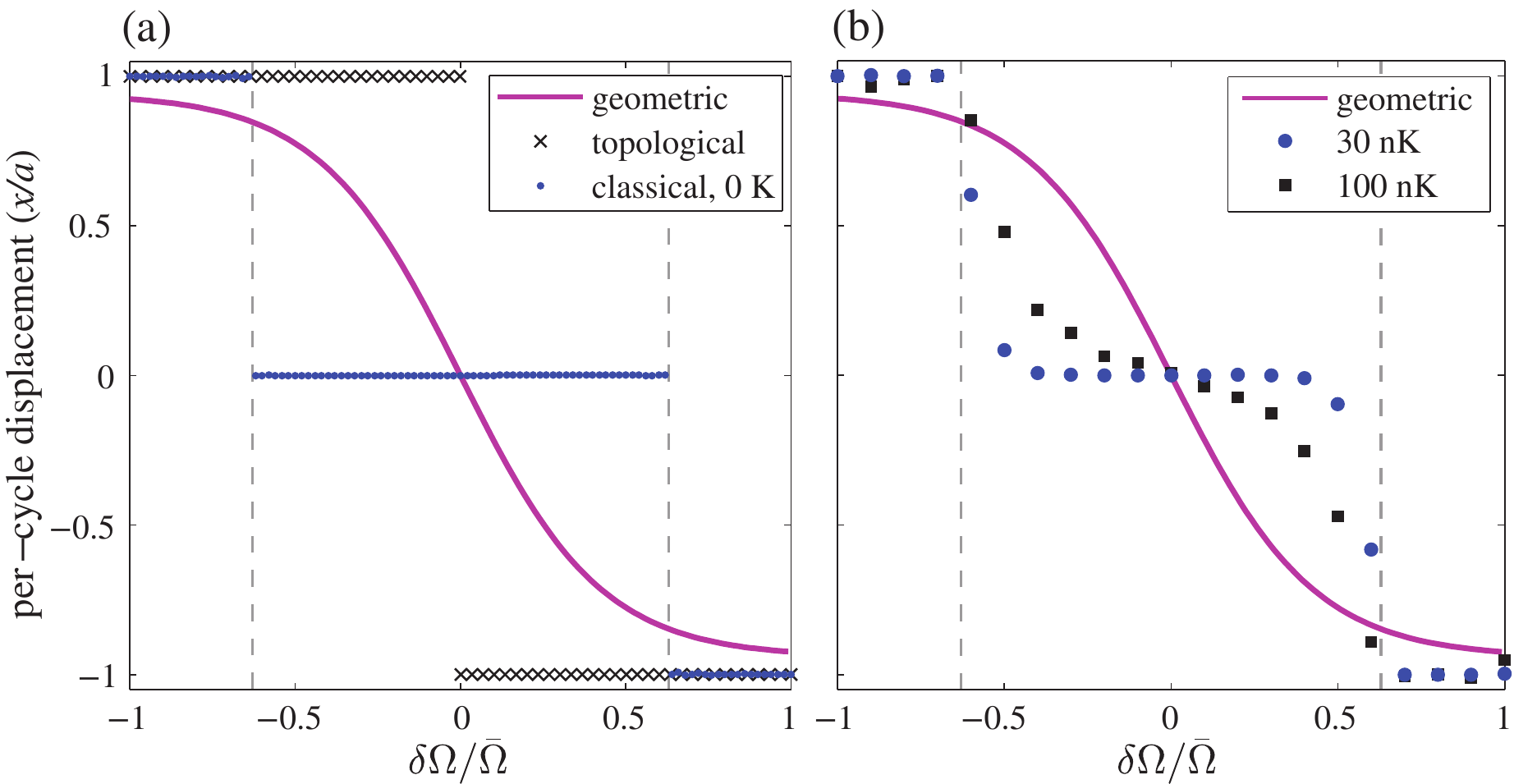}
    \caption{(a) Computed per-cycle displacement versus Raman imbalance $\delta\Omega/\bar\Omega$ for geometric, topological, and classical pumps. Classical particles have zero initial kinetic energy. Grey dashed lines indicate the critical Raman imbalance values $\delta\Omega/\bar\Omega=\pm 0.63$. (b) Computed per-cycle displacement versus $\delta\Omega/\bar\Omega$ for classical ensembles at finite temperatures (30 nK and 100 nK) is shown in symbols. Ramping period for classical pumps is 2 ms.}
    \label{ThreePump}
\end{figure*}
Our BECs had a typical temperature of $20$ nK with $\approx 5\%$ of the atoms in the thermal component.
To show the contrast between a geometric pump and a classical pump, we first simulated trajectories of particles at an initial temperature of 0 K, which is the closest condition for describing the majority of atoms occupying the lowest energy state of the lattice. The classical trajectory is governed by the force derived from the lattice potential. Figure~\ref{ThreePump}(a) shows the simulated per-cycle displacement versus $\delta\Omega/\bar\Omega$ for a geometric pump and a classical pump. We also included the topological pump result, which was obtained by integrating the Zak phase [Fig.~3(a) in the main text] over rf phase, describing the displacement of a filled band. The topological pump displays quantized per-cycle displacement as expected; the sign of the displacement depends on the sign of Raman imbalance, since the Zak phase in Fig.~3(a) of the main text also has mirror symmetry across zero Raman imbalance.

The classical pump also displays quantized displacement. The transition point from zero to non-zero per-cycle displacement occurs at $\delta\Omega/\bar\Omega\approx \pm 0.63$, which is consistent with the underlying motion of the adiabatic potential [Fig.~3(c) in the main text]. For large Raman imbalances, the adiabatic potential is displaced by one site per cycle, shifting the classical particle by one site as well; for small imbalances, two sublattices always maintain local energy minima, giving no net force on the classical particle released from local energy minima.

We also computed the classical pump for atoms at non-zero temperatures: 30 nK and 100 nK. The initial thermal distribution was taken to occupy a single sublattice centered at the origin (which has the lower energy at $\phi=0$ compared to the other sublattice regardless of the Raman imbalance), and the resulting per-cycle displacements are shown in Fig.~\ref{ThreePump}(b). The 30 nK system is nearly quantized over a wide range of Raman imbalances, while the 100 nK ensemble deviates by more from the quantized values. In either case, the qualitative dependence on the Raman imbalance is in stark contrast to the geometric pump (solid line).

\section{Berry curvature at non-zero crystal momentum}
\begin{figure*}[]
  \centering
    \includegraphics[width=5in]{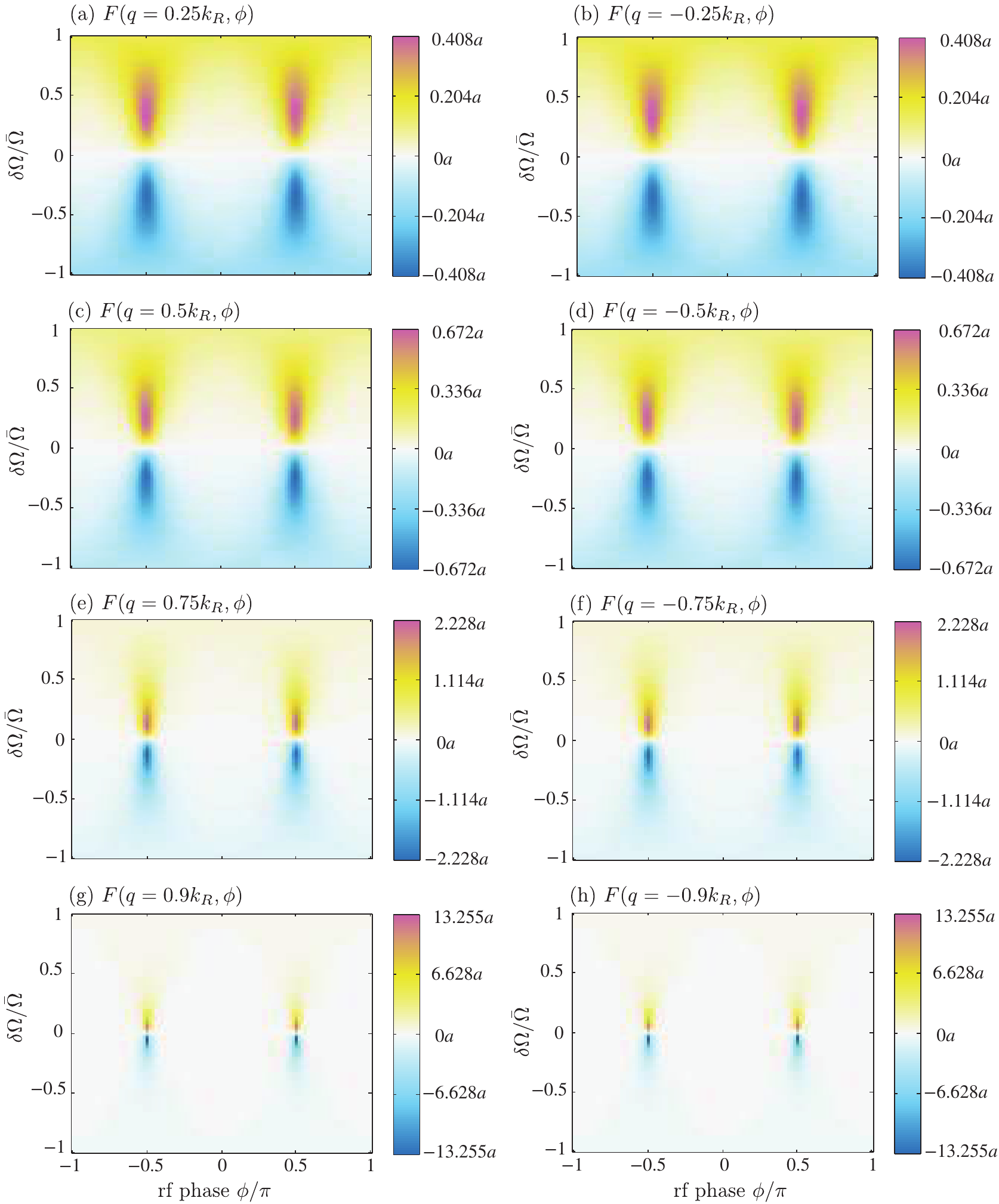}
    \caption{Berry curvature $F(q,\phi)$ versus Raman imbalance $\delta\Omega/\bar\Omega$ and rf phase $\phi/\pi$. (a-b): $q=\pm0.25k_R$; (c-d): $q=\pm0.5k_R$; (e-f): $q=\pm0.75k_R$; (g-h): $q=\pm0.9k_R$. $\hbar({\bar{\Omega},\Omega_{\rm rf}},\delta)=(6,2.2,0)E_R$.}
    \label{BerryCruvature_qs}
\end{figure*}

\begin{figure}[h!]
  \centering
    \includegraphics[width=3in]{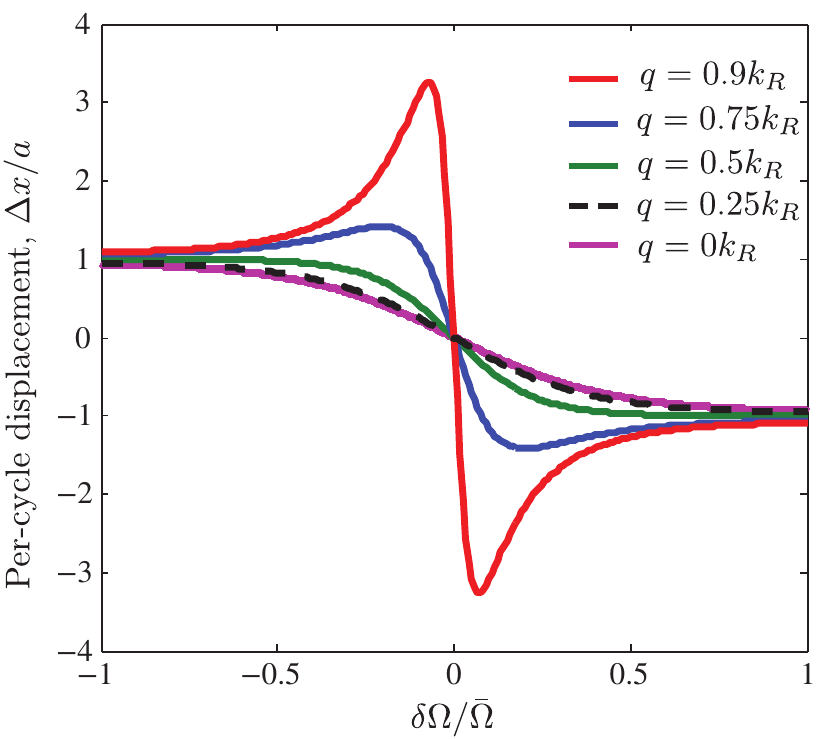}
    \caption{Per-cycle displacement versus Raman imbalance for different crystal momentum values. Per-cycle displacements were obtained by integrating Berry curvatures in Fig.~\ref{BerryCruvature_qs}(a,c,e,g) over the rf phase.}
    \label{Displacement}
\end{figure}

\begin{figure*}[]
  \centering
    \includegraphics[width=5in]{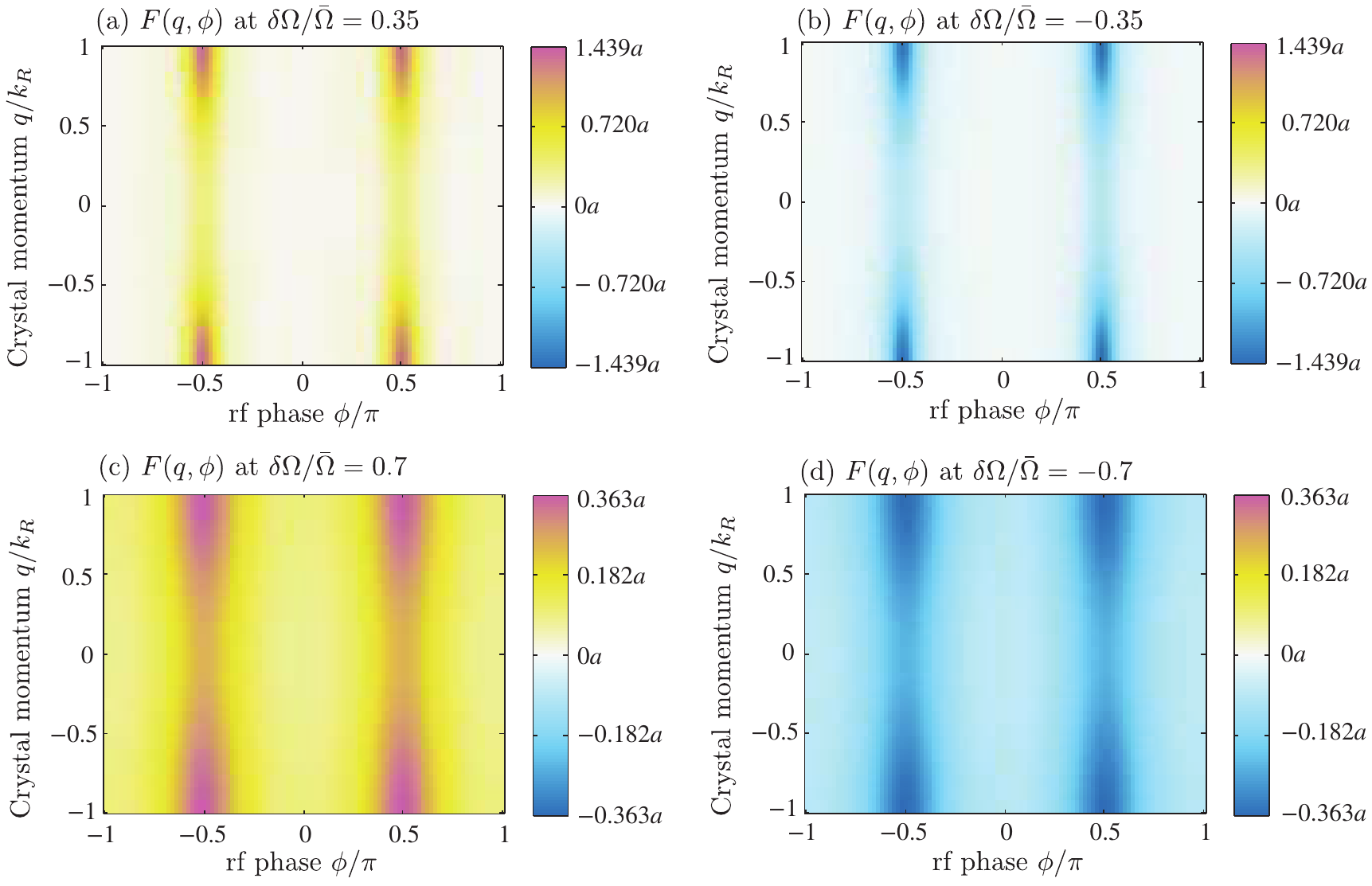}
    \caption{Dependance of Berry curvature on $q$ and $\phi$ for four Raman imbalances. $\hbar({\bar{\Omega},\Omega_{\rm rf}},\delta)=(6,2.2,0)E_R$.}
    \label{BerryCruvature_q_dep}
\end{figure*}

\begin{figure*}[]
  \centering
    \includegraphics[width=5.0in]{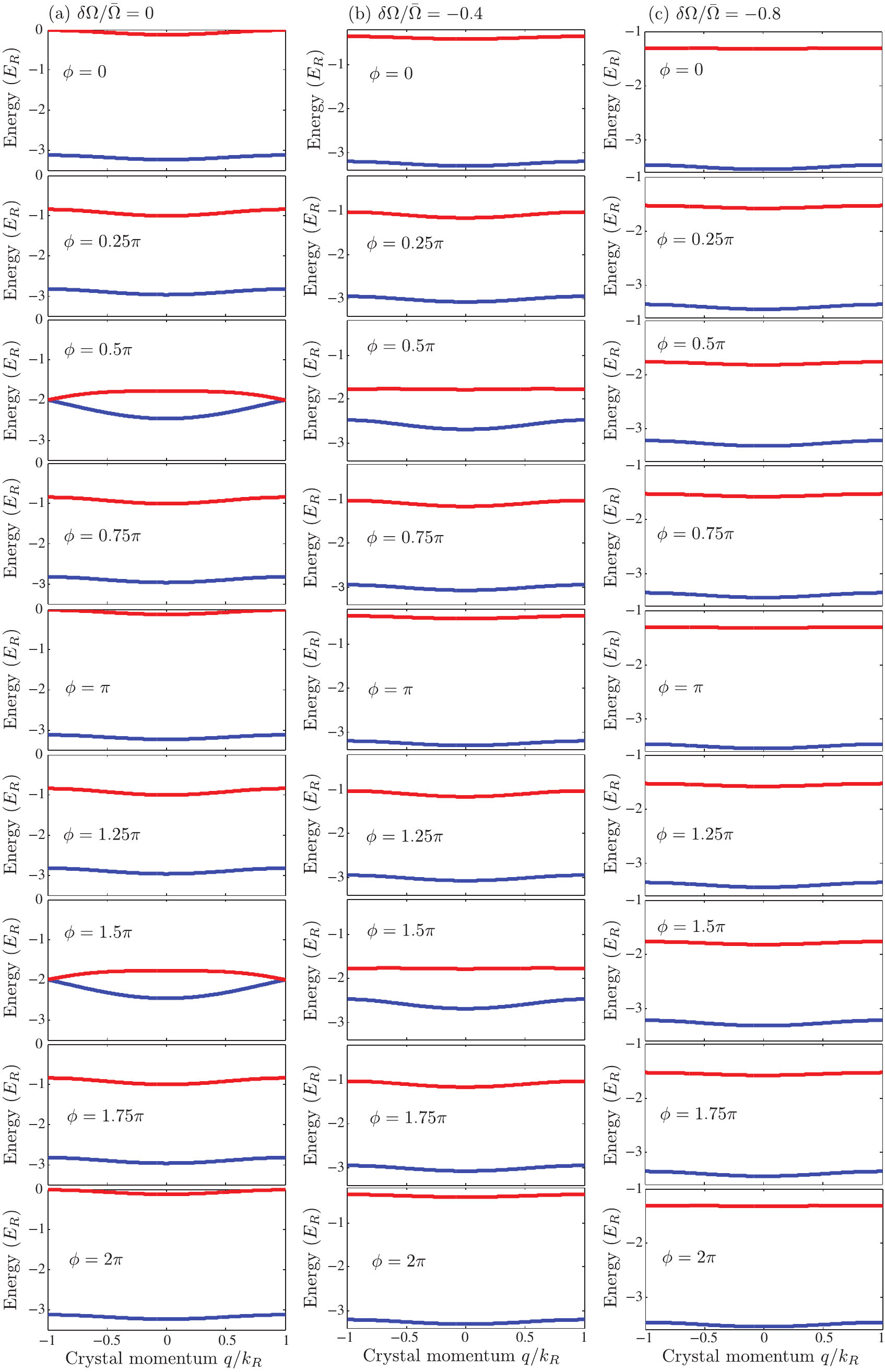}
    \caption{Band structures of the lowest two bands of the magnetic lattice at three Raman imbalances: (a) $\delta\Omega/\bar\Omega=0$, (b) $\delta\Omega/\bar\Omega=-0.4$, and (c) $\delta\Omega/\bar\Omega=-0.8$. For each Raman imbalance, we plot a series of rf phases $\phi$ (from top to down) consisting of a pump cycle. $\hbar({\bar{\Omega},\Omega_{\rm rf}},\delta)=(6,2.2,0)E_R$.}
    \label{BandStructure}
\end{figure*}

In our charge pump experiment, we prepared the BEC at $q=0$ with a small momentum width of $0.004k_R$. Therefore, we only consider the Berry curvature $F(q=0,\phi)$ in the main text. BECs in a lattice at different crystal momenta can be prepared experimentally using Bloch oscillation \cite{Dahan1996} (such as applying optical potential gradient), prompting us to investigate the Berry curvature of our lattice at $q\neq0$. Figure~\ref{BerryCruvature_qs} plots the Berry curvature $F(q,\phi)$ at several $q$ values versus the two experimental parameters, Raman imbalance $\delta\Omega/\bar\Omega$ and rf phase $\phi$. To evaluate the Berry curvature, we used the discrete approach [18]. First, we defined the phase angle $\gamma$ between two states as $\exp(-i\gamma)=\langle u(R)|u(R+\delta R)\rangle=1+\delta R \langle u|\partial_R u \rangle$, where $R=q$ or $\phi$. By taking the logarithm of the previous equation, we obtained $\gamma=-$\rm{Im log}$\langle u(R)|u(R+\delta R)\rangle=\delta R i \langle u|\partial_R u \rangle$. Therefore, we can evaluate $F(q,\phi)$ as follows:

\begin{align}
F(q,\phi)&=i(\langle \partial_q u| \partial_\phi u\rangle- \langle\partial_\phi u|\partial_q u\rangle)  \nonumber \\
&=i \frac{\partial}{\partial q} \langle u| \partial_\phi u\rangle-i \frac{\partial}{\partial \phi} \langle u| \partial_q u\rangle \nonumber \\
&=\frac{1}{\delta q \delta \phi}[-\rm{Im \,log} \langle u_2|u_4\rangle+\rm{Im \,log} \langle u_1|u_3\rangle] \nonumber \\
&-\frac{1}{\delta \phi \delta q }[-\rm{Im \, log} \langle u_3|u_4\rangle+\rm{Im \, log} \langle u_1|u_2\rangle],
\label{eq:F_q_phi}
\end{align}
where we define $\ket{u_1}=\ket{u(q,\phi)}, \ket{u_2}=\ket{u(q+\delta q,\phi)}, \ket{u_3}=\ket{u(q,\phi+\delta\phi)}$, and $\ket{u_4}=\ket{u(q+\delta q,\phi+\delta \phi)}$; we used $\delta q=10^{-3}k_R$ and $\delta \phi=10^{-3}$ for computation. When $q$ approaches the end of BZ, the largest Berry curvature value increases and $F(q,\phi)$ becomes highly concentrated near points with zero Raman imbalance and $\phi=(2n+1)\times \pi/2$. Band gaps close at these points in the end of BZ, as shown in Fig.~\ref{BandStructure}(a). Topological phase transition occurs when these points are traversed where Zak phases possess singularities [Fig.~3(a) in the main text]. In addition, Berry curvatures are even in $q$, i.e. $F(q,\phi)=F(-q,\phi)$. We want to show next that $F(q,\phi)$ even in $q$ is a general consequence of a 1D lattice possessing a time-reversal symmetry. Note that time-reversal symmetry operates on the "frozen" lattice at any moment during pumping (lattice Hamiltonian during a pump cycle does not satisfies $H(-t)=H(-t)$, which would give zero displacement). The Bloch states at $q$ and $-q$ can at most differ by a phase, i.e. $\hat T u_{q,\phi}(x)=u_{-q,\phi}(x)^*=\exp(i\beta(q,\phi))u_{q,\phi}(x)$, where $\hat T$ is the time reversal operator. Using this gauge and Eq.~\ref{eq:F_q_phi}, we can show $F(-q,\phi)=F(q,\phi)+1/(\delta \phi \delta q)[-(\beta_4-\beta_2)+(\beta_3-\beta_1)+(\beta_4-\beta_3)-(\beta_2-\beta_1)]=F(q,\phi)$. Here, $\exp(i\beta_j)$ is gauge term for state $|u_j\rangle$. This yields Berry curvatures which are even in $q$ as well.

The per-cycle displacement at different $q$ values can be obtained by integrating the Berry curvature over the rf phase in a pump cycle $\Delta x(q)=-\int F(q,\phi) d\phi$, resulting in Fig.~\ref{Displacement}. Since the Berry curvature is even in $q$, we only showed the displacement for positive $q$. The per-cycle displacements versus Raman imbalance are similar between $q=0$ and $q=0.25 k_R$: when reducing the imbalance value from $|\delta\Omega/\bar\Omega|=1\rightarrow 0$, the displacement value gradually reduces from less than $|a|$ to zero. At $q=0.5k_R$, the displacements reach $\pm a$ at large Raman imbalances. For $q>0.5k_R$, the displacement demonstrates a significantly different dependence on Raman imbalance:  displacement value increases first when reducing the imbalance value. The displacement value is much larger than $|a|$ when $|q|$ approaches $k_R$. This reflects the large and highly concentrated Berry curvatures near zero Raman imbalance, shown in Fig.~\ref{BerryCruvature_qs}(g-f).

When averaging the displacements of the geometric pump results for the entire BZ, we should obtain the result of a topological pump, as shown as black crosses in Fig.~\ref{ThreePump}(a). Since the topological pump yields quantized displacement $\pm a$ depending only on the sign of Raman imbalance, the non-classically large displacement at low Raman imbalance and large $q$ in Fig.~\ref{Displacement} compensates the small displacement at low $q$.
In Fig.~\ref{BerryCruvature_q_dep}, we showed the Berry curvature $F(q,\phi)$ on the plane spanned by crystal momentum $q$ and rf phase $\phi$, which effectively forms a torus for a filled band during charge pumping. Berry curvatures are concentrated toward $q=\pm k_R$ at $\phi=\pm \pi/2$, which is more obvious for low Raman imbalance [Fig.~\ref{BerryCruvature_q_dep}(a-b)]. In addition, low Raman imbalance has a larger $|F(q,\phi)|_{max}$ than high Raman imbalance. We integrated $F(q,\phi)$ in Fig.~\ref{BerryCruvature_q_dep} to obtain the per-cycle displacement and indeed obtained quantized displacement of $+a (-a)$ for $\delta\Omega/\bar\Omega<0$ ($\delta\Omega/\bar\Omega>0$), which is consistent with the topological pump result [black crosses in Fig.~\ref{ThreePump}(a)].

\section{Band structures}

The band structures of the lowest two bands are shown in Fig.~\ref{BandStructure} for three Raman imbalances. For each imbalance, we showed the band structures as a function of rf phase in a pump cycle (from top to down). We first noticed that the band structure is even in crystal momentum at any $\phi$, i.e. $\epsilon(q)=\epsilon(-q)$. In general, a 1D lattice Hamiltonian preserving time-reversal symmetry, such as the SSH model, yields a band structure which is symmetric in $q$.

From the above two properties, our pumping scheme can be extended to probe Berry curvatures at $q\neq 0$ from a pairwise measurement. Using Bloch oscillations [36], we can prepare BECs at any $\pm |q_0|$ and ramp the rf phase as we showed in the main text. Since the dynamic phases from $\pm |q_0|$ are opposite (symmetric band structure in $q$ gives opposite group velocities) but the geometric phase contributions are the same, the sum of the per-cycle displacements at $\pm |q_0|$ only probes the Berry curvature at $q_0$. Specifically, $\Delta x(+|q_0|)+\Delta x(-|q_0|)=-\int F(+|q_0|,\phi) d\phi-\int F(-|q_0|,\phi) d\phi=-2\int F(|q_0|,\phi) d\phi$. After repeating the measurement for the several $q$ values, one can obtain per-cycle displacements shown in Fig.~\ref{Displacement}, and hence map out the geometry of the entire band.

\section{Fitting procedure to obtain per-cycle displacement and State Preparation}
\begin{figure}[h!]
  \centering
    \includegraphics[width=3.0in]{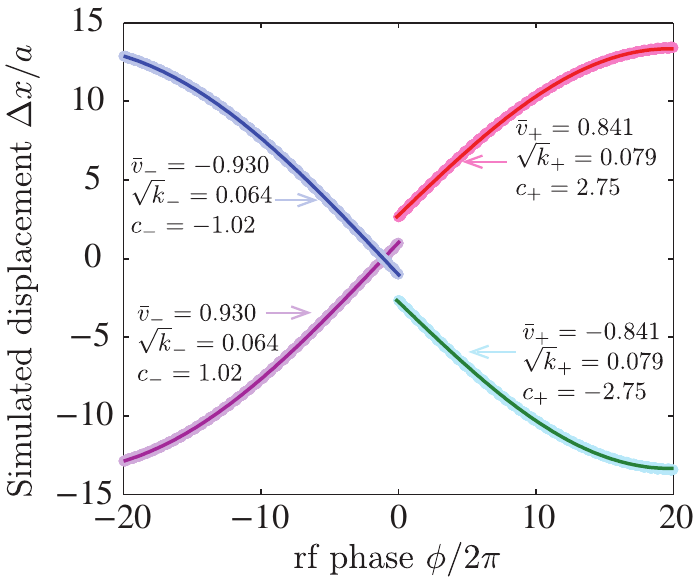}
    \caption{Simulated displacements (thick lines) in charge pumping for Raman imbalance $\delta\Omega/\bar \Omega= \pm0.7$. $\hbar({\bar{\Omega}, \Omega_{\rm rf}},\delta)=(6.38,2.2,0)E_R$. Thin lines show the fitting results of the simulated displacement using Eq.~\ref{eq:fit_fun}.}
    \label{Simulated_displacement}
\end{figure}

We expected the displacement to have a linear dependence on pump cycles from the geometric pumping with a per-cycle displacement given by $\Delta x(q=0)=\bar v\times T=-\int F(q,\phi) d\phi$. However, the restoring force from the dipole trap would impact the actual displacements. We performed real-space simulations of the charge pumping process using the split operator method in the presence of a harmonic trapping potential, as shown in Fig.~\ref{Simulated_displacement} for Raman imbalance $\pm0.7$ (thick lines). To extract the contribution of geometric pumping, we can describe the simulated displacements as
\begin{equation}
\Delta x_{\pm}=\bar v_{\pm}/\sqrt k_{\pm}\sin(\sqrt k_{\pm} \times \phi)+c_{\pm},
\label{eq:fit_fun}
\end{equation}
where subscripts $\pm$ represent the directions of increasing and decreasing phases. Here, $k_{\pm}$ are the effective spring constants from the trapping potential and magnetic lattice potential.

During charge pumping, the phase was linearly ramped $\phi = 2\pi t/T$.
To linearly ramp the rf phase, we set a small frequency difference between the rf and Raman coupling fields as described before. Due to technical constraint on the DDS signal generator, we had to introduce the frequency difference in the beginning of the state preparation. During state preparation, we ramped the coupling fields up to constant values and an initial detuning ($\delta_i=2.5E_R$) down to zero. We found through simulations that ramping the rf phase during state preparation introduced small offsets $c_{\pm}$ (Note that rf phases remained at constant values and $\delta_i=5E_R$ during state preparation for Fig.~2 of the main text).
Thin lines in Fig.~\ref{Simulated_displacement} show the fitting results using Eq.~\ref{eq:fit_fun}. The difference between the average value of $\bar v_{\pm}$ from the fits and the expected per-cycle displacement from integrating Berry curvature is $<1\%$.

We can further fit our data at non-zero Raman imbalances [such as symbols in Fig.~4(b) of the main text] to the functional form:
$\Delta x= \bar v_{+,\rm{exp}}/\sqrt k_{+}\sin(\sqrt k_{+} \times \phi)+c_{+}+C$ for $\phi>0$ and
$\Delta x=\bar v_{-,\rm{exp}}/\sqrt k_{-}\sin(\sqrt k_{-} \times \phi)+c_{-}+C$ for $\phi<0$, where the three fitting parameters are $\bar v_{\pm,\rm{exp}}$ and an overall offset $C$; the rest of the parameters are taken from the fit results to the simulated displacements. The mean of $\bar v_{\pm,\rm{exp}}$ extracts the per-cycle displacement from our data.
For zero Raman imbalance, we simply fitted a linear line to the data shown as black symbols in Fig.~4(b) of the main text.

\section{Field stabilization}
We took two in-situ partial-transfer images using two microwave pulses which were $\pm$500 Hz detuned from $\omega_Z$ in all of our measurements. Based on the asymmetry (ratio of difference to sum) of two atomic signals, we actively stabilized the laboratory bias magnetic field after each measurement. This can only correct a long term field drift in the laboratory. We rejected the data point if the asymmetry was larger than 0.25. Based on our independent field calibrations from pulsing a single Raman or rf coupling field alone, the post-selected atoms experienced a detuning which was within the quadratic Zeeman shift of $\hbar \epsilon\approx h\times100$ Hz.
We intentionally introduced some detuning $\delta_i$ (by changing the bias field by a known amount) and then ramped it down during the state preparation, which can adiabatically load the BECs into the ground state of the magnetic lattice.

For Raman imbalance of $\pm$0.7 [trajectory \rmnum{1}~and \rmnum{3}~in Fig. 4(b) of the main text], we took 30 images for each data point; for zero imbalance [trajectory \rmnum{2}~in Fig. 4(b)] and imbalance of $\pm$0.35 (trajectory not shown) we took 10 images for each data point. Since we rejected data points based on the partial-transfer atom signals, we ended up keeping $60\%-70\%$ of the images for each data point.

\section{Adiabaticity of charge pump}
As shown in Fig.~\ref{BandStructure}, the smallest gap at $q=0$ is $\approx 1E_R$ occurring at zero imbalance and $\phi=(2n+1)\times \pi/2$. Our pumping speed of 500 Hz for non-zero Raman imbalance is appreciably slower than $E_R/h=3.65$ kHz, ensuring the adiabaticity. In addition, we had tried a factor of two larger ramping speed than 500 Hz in the experiment at Raman imbalance of $+0.7$ and observed no significant difference in both the displacement and magnetization measurements.

After each measurement of displacement, we also recorded the time of flight image of the same BEC. The contrast of the diffraction pattern (showing the distribution in both momentum and spin) remained sharp during our pumping experiment: a good added indication of an adiabatic experiment. In addition, the magnetization measurement shown in the manuscript displays a good contrast for more than 50 pump cycles at the ramping rate we chose, confirming again the adiabaticity of our pumping process.

\end{document}